\documentclass{iopart}

\usepackage{iopams}
\usepackage{url}
\usepackage[ngerman,american]{babel}
\usepackage{pstricks}
\usepackage{pst-plot}
\usepackage{pst-all}

\newcommand{\D}{\mathrm{d}}

\newcommand{\diff}[2]{\frac{\D #1}{\D #2}}
\newcommand{\piff}[2]{\frac{\partial #1}{\partial #2}}

\begin{document}

\title{Brachistochrones With Loose Ends}

\author{Stephan Mertens$^{1,2}$ and Sebastian Mingramm$^1$}

\address{$^1$ \selectlanguage{ngerman}{Institut\ f"ur\ Theoretische\ Physik,
    Otto-von-Guericke Universit"at, PF~4120, 39016 Magdeburg, Germany}} 

\address{$^2$ Santa Fe Institute, 1399 Hyde Park Road,
Santa Fe, New Mexico 87501, U.S.A.}

\date{\today}

\begin{abstract}
  The classical problem of the brachistochrone asks for the curve down
  which a body sliding from rest and accelerated by gravity will slip
  (without friction) from one point to another in least time.  In
  undergraduate courses on classical mechanics, the solution of this
  problem is the primary example of the power of the variational
  calculus. Here we address the generalized brachistochrone problem
  that asks for the fastest sliding curve between a point and a given
  curve or between two given curves. The generalized problem can be
  solved by considering variations with varying endpoints.  We will
  contrast the formal solution with a much simpler solution based on
  symmetry and kinematic reasoning. Our exposition should encourage
  teachers to include variational problems with free boundary
  conditions in their courses and students to try simple, intuitive
  solutions first.
\end{abstract}


% 45.10.Db 	Variational and optimization methods
% 45.50.Dd 	General motion
% 02.30.Xx 	Calculus of variations

%\maketitle

\section{Introduction}

The Brachistochrone Transit Company BTC \cite{calvert:BTC}, well known
for its highspeed, gravity-propelled point-to-point transportation
services, wanted to enter the coal chute market. Coal chutes are
usually used to overcome a given horizontal distance $\Delta x$, the
vertical distance $\Delta y$ being irrelevant. The chief engineer of
BTC, Dr.~L.A.~Grange, faced the problem of designing the optimal
chute.  Dr.~Grange immediately realized that the optimal coal chute is
given by one of the companies most successful sliding devices, the
\textsc{Cycloid}\texttrademark. His argument:
\begin{itemize}
\item[(A)] The optimal coal chute will run between two points $(x,y)$
  and $(x+\Delta x,y+\Delta y)$.  But the fastest sliding curve
  between two given points is a cycloid.
\end{itemize}
The problem then is to select the \emph{best} cycloid or, equivalently,
the corresponding vertical distance $\Delta y$.

\begin{figure}
 \begin{center}
 \begin{pspicture}(0,0)(6.5,-6)
   %\psgrid
   \psset{arrowsize=2pt 4}
   %\qdisk(0,0){4pt}
   \psset{linecolor=black}
   \pnode(6.5,0){optistart}
   \pnode(6.5,-3.82){optiziel}
   \ncline{<->}{optistart}{optiziel}
   \ncput*{$\displaystyle \frac{2}{\pi}\Delta x$}
   \pnode(0,0){start}
   \pnode(6,0){ziel}
   \ncline{<->}{start}{ziel}
   \ncput*{$\Delta x$}
   \pnode(6,-6){lineziel}
   \ncline{<->}{ziel}{lineziel}
   \ncput*{$\Delta x$}

   %\rput(6.5,-3.82){$\displaystyle \frac{2}{\pi}\Delta x$}
   \psline[linestyle=solid](0,0)(6,-6)
   \parametricplot[plotstyle=curve,linestyle=dashed,linecolor=black]{0}{2.412}{t t 57.2958 neg mul sin add 6 mul 1.745453416 div t 57.2958 mul cos 1 sub 6 mul 1.745453416 div}
   \parametricplot[plotstyle=curve,linestyle=solid,linecolor=black]{0}{3.1415}{t t 57.2958 neg mul sin add 6 mul 3.1415 div t 57.2958 mul cos 1 sub 6 mul 3.1415 div}
 \end{pspicture}
 \caption{\label{fig:coal-chutes} Covering a horizontal distance $\Delta x$ by sliding
 down the fastest straight line, a cycloid with the same endpoint (dashed) or the optimal cycloid (solid).}
 \end{center}
\end{figure}
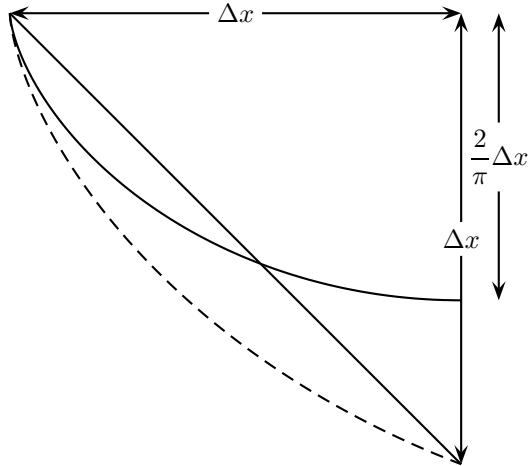

Coal chutes that are used in practice are simple inclined planes.
Under the usual assumptions (no friction, start from rest, homogeneous
gravitational acceleration $g$), the time to cover a horizontal
distance $\Delta x$ by sliding down a straight line with inclination
$\alpha$ reads
\begin{equation}
  \label{eq:T-line}
  T_s(\alpha) = \sqrt{\frac{\Delta x}{g}}\,\sqrt{\frac{2}{\sin\alpha \cos\alpha}}\,.
\end{equation}
This time is minimal for $\alpha=\frac{\pi}{4}$, i.e., for a line that spans the same horizontal and vertical distance. The minimal time is 
\begin{equation}
  \label{eq:T-opt-line}
  T_s^\star = T_s(\frac{\pi}{4}) = 2\,\sqrt{\frac{\Delta x}{g}}\,.
\end{equation}
A (downward) cycloid with starting point $(0,0)$ is
parameterized by
\begin{equation}
  \label{eq:cycloid}
    x(t) = a(t-\sin t) \qquad
    y(t) = -a(1-\cos t)
  \qquad 0\leq t\leq t_0\,,
\end{equation}
where the constants $a$ and $t_0$ are determined by the end point
$(\Delta x,\Delta y)$.  The time to slide down a cycloid to $(\Delta
x, \Delta x)$ is
\begin{equation}
  \label{eq:T-cycloid}
  T_c =  \frac{t_0}{\sqrt{1-\cos t_0}}\, \sqrt{\frac{\Delta x}{g}} =
  1.826 \,  \sqrt{\frac{\Delta x}{g}}
\end{equation}
where $t_0=2.412\ldots$ is the positive root of $t-\sin t = 1-\cos t$.
This is faster than along the straight line, but is it the fastest
way? Dr.\ Grange said no, and he gave the following argument: 
\begin{itemize}
\item[(B)] Suppose we are sliding down the optimal curve.  When we are
  very close to the vertical line at $\Delta x$, we can assume that we
  will reach it in the next infinitesimal time step $\D t$. For this
  infinitesimal time step we can ignore the acceleration and assume
  that we will continue with constant velocity. But the fastest way to
  the vertical line is to move horizontally, i.e., to \emph{meet the
  target curve under a right angle.}
\end{itemize}
Dr.~Grange quickly calculated that the cycloid that intersect the
vertical line at $\Delta x$ under a right angle is the one with end
point $(\Delta x,\frac{2}{\pi}\Delta x)$ (Fig.~\ref{fig:coal-chutes}),
and the time to slide down this cycloid reads
\begin{equation}
  \label{eq:T-opt-cycloid}
  T_c^\star =  \sqrt{\pi}\, \sqrt{\frac{\Delta x}{g}} =
  1.772 \, \sqrt{\frac{\Delta x}{g}}\,.
\end{equation}
This is in fact the optimum as Dr.~Grange verified by trying nearby
cycloids. Dr.~Grange reported the solution to his boss, and BTC became
the major player in the coal chute industry.

\section{Variations with varying ends}
\label{sec:varying-end}

The story of Dr.~Grange and BTC illustrates two facts: (1) the problem
of the baristochrone with varying endpoint(s) is largely unknown, and
(2) its solution requires only a simple argument in addition to the
solution of the classical, fixed endpoint version.  In his 1696 paper 
\cite{bernoulli:1696},
Johann Bernoulli challenged the learned world with the fixed endpoint
problem (English translation from \cite{struik:69}):
\begin{quote}
  Let two points $A$ and $B$ given in a vertical plane. To find the
  curve that point $M$, moving on a path $AMB$, must follow that,
  starting from $A$, it reaches $B$ in the shortest time under its own
  gravity.
\end{quote}
It is this version that dominates modern textbooks.  The more general
version of the problem, where $A$ or $B$ (or both) are free to move
along given curves, is rarely discussed. Notable exceptions are
Gelfand and Fomin \cite{gelfand:fomin} or Weinstock \cite{weinstock}.
This is astonishing considered the fact that the generalized problem
was already solved by Joseph Louis Lagrange in 1760 
\cite{lagrange:1760}, in the same paper that laid the foundation of the variational
calculus. Before we present the formal solution let us reconsider
Dr.~Grange's arguments. Obviously his argument (A) is generally true,
i.e., the brachistochrone is always found among the solutions of the
corresponding Euler differential equation. The endpoints
(fixed or free) are only used to fix the the free parameters of the
solution. His argument (B) is also much more general than suggested by
the coal chute example. It is a kinematic argument that applies for
\emph{any} gravitional field and \emph{any} target curve, not just
the vertical line: a brachistochrone will intersect its target curve
orthogonally. Let us see whether this is confirmed by the
variational calculus.

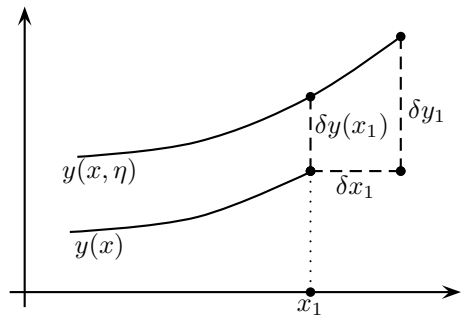
\begin{figure}
 \begin{center}
 \begin{pspicture}(0,0)(6,4)
   %\psgrid
   \psset{arrowsize=2pt 4}
   \psset{linecolor=black}
   \psline{->}(0,0.2)(6,0.2)
   \psline{->}(0.2,0)(0.2,4)
   \listplot[plotstyle=curve,showpoints=false]{0.8 1 2.5 1.2 4 1.8} 
   \listplot[plotstyle=curve,showpoints=false]{0.9 2 2.5 2.2 4 2.8 5.2 3.6}  
   \listplot[plotstyle=line,linestyle=dashed,showpoints=true]{4 1.81 4 2.8}
   \listplot[plotstyle=line,linestyle=dashed,showpoints=true]{4 1.81 5.2 1.81}
   \listplot[plotstyle=line,linestyle=dashed,showpoints=true]{5.2 1.81 5.2 3.6}
   \rput(1.2,0.8){$y(x)$}
   \rput(1.2,1.8){$y(x,\eta)$}
   \rput(4.55,2.4){$\delta y(x_1)$}
   \rput(5.55,2.6){$\delta y_1$}
   \rput(4.6,1.6){$\delta x_1$}
   \listplot[plotstyle=line,linestyle=dotted,showpoints=true]{4 0.2 4 1.81}
   \rput(4,0){$x_1$}
 \end{pspicture}
 \caption{\label{fig:endpoints} Variation of a curve $y(x)$ and its endpoint.}
 \end{center}
\end{figure}

We want to solve the following variatonal problem: Find the minimum of
\begin{equation}
  \label{eq:def-problem}
  J[y] = \int\limits_{x_0}^{x_1} F(x,y,y')\,\D x
\end{equation}
with respect to variations of $y$ and variations of the endpoints
$(x_0,y_0)$, $(x_1,y_1)$. To find the extremum of the functional
$J[y]$, we consider a family of functions $y(x,\eta)$ such that
$y(x,0) = y(x)$. The variations of $y$ and $y'$ are defined as usual
by
\begin{displaymath}
  \delta y = \left.\piff{y(x,\eta)}{\eta}\right|_{\eta=0}\cdot\eta
\end{displaymath}
and
\begin{displaymath}
  \delta y'= \left.\piff{}{\eta}\left[\piff{y(x,\eta)}{x}\right]\right|_{\eta=0}\cdot\eta
  = \diff{}{x} \delta y\,.
\end{displaymath}
We also vary the endpoints $\big(x_i(\eta),y_i(\eta)\big)$ such that
$\big(x_i(0),y_i(0)\big)=(x_i,y_i)$ for $i=0,1$. We only need to
consider variations of the $x$-coordinates,
\begin{displaymath}
  \delta x_i = \left.\diff{x_i(\eta)}{\eta}\right|_{\eta=0}\cdot\eta\,,
\end{displaymath}
because the variations of $y$-coordinates are given by $\delta x_i$ and the
variation of the curve $y(x)$ (Fig.~\ref{fig:endpoints}):
\begin{equation}
  \label{eq:delta-y}
  \delta y_i = \delta y(x_i) + y'(x_i)\,\delta x_i\,.
\end{equation}
For this family of functions and endpoints, $J[y]$ becomes a function
of $\eta$,
\begin{equation}
  \label{eq:J-eta}
  J(\eta) = \int\limits_{x_0(\eta)}^{x_1(\eta)} F\big(x,y(x,\eta),y'(x,\eta)\big)\,\D x\,.
\end{equation}
Differentiating $J(\eta)$ at $\eta=0$ and multiplying by $\eta$ yields
the first variation of $J$:
\begin{equation}
  \label{eq:first-variation-1}
  \delta J = \big[ F\,\delta x\big]_0^1 + \int\limits_{x_0}^{x_1}\left(F_y\delta y + F_{y'}\delta y'\right) \D x\,,
\end{equation}
where we have used the shorthands $F_y=\partial F/\partial y$ etc.\
and
\begin{displaymath}
  \big[ F\,\delta x\big]_0^1 = F(x_1,y(x_1),y'(x_1))\,\delta x_1 - F(x_0,y(x_0),y'(x_0))\,\delta x_0\,.
\end{displaymath}
Partial integration of the second term in the integral in
\eref{eq:first-variation-1} yields
\begin{displaymath}
  \int\limits_{x_0}^{x_1} F_{y'}\diff{}{x}\delta y\,\D x = \left[F_{y'} \delta y\right]_{x_0}^{x_1} - \int\limits_{x_0}^{x_1} \left(\diff{}{x}F_{y'}\right) \delta y\,\D x\,.
\end{displaymath}
This and \eref{eq:delta-y} allow us to write the variation of $J$ as
\begin{equation}
  \label{eq:first-variation-2}
  \delta J = \left[ \big(F - y'F_{y'}\big)\delta x + F_{y'}\,\delta y\right]_0^1 
   + \int\limits_{x_0}^{x_1}\left(F_y -\diff{}{x} F_{y'}\right)\delta y\, \D x\,.
\end{equation}
A minimum of $J$ implies $\delta J = 0$. In particular $\delta J=0$
for the subset of variations which leave the endpoints fixed, i.e., for $\delta
x_i=0$ and $\delta y_i=0$.  This implies that the integrand in
\eref{eq:first-variation-2} must vanish, which brings us to the Euler
equation
\begin{equation}
  \label{eq:euler}
  F_y -\diff{}{x} F_{y'} = 0\,.
\end{equation}
This proves Dr.~Grange's claim (A) that the optimal curves for varying
endpoints are from the same family of curves that are optimal for
fixed endpoints.  If we allow varying endpoints as well, $\delta J=0$
in addition implies
\begin{equation}
  \label{eq:vary-endpoints}
  \left.(F-y'F_{y'})\right|_{x=x_i}\delta x_i + \left.F_{y'}\right|_{x=x_i}\delta y_i = 0 \qquad i=0,1\,.
\end{equation}
Now we consider variations where the endpoints lie on given curves
$g_0(x_0,y_0)=0$ and $g_1(x_1,y_1)=0$. This gives us another equation
for $\delta x_i$ and $\delta y_i$, namely
\begin{equation}
   \label{eq:endpoints-on-curves}
   \piff{g_i(x_i,y_i)}{x} \delta x_i + \piff{g_i(x_i,y_i)}{y} \delta y_i = 0
   \qquad i = 0,1\,.
\end{equation}
For a non-trivial solution to exist, the determinant of the coeffcients of
the system \eref{eq:vary-endpoints} and \eref{eq:endpoints-on-curves}
has to vanish, i.e., we require
\begin{equation}
  \label{eq:transversality}
  \left.\partial_xg_i F_{y'}\right|_{x=x_i} = \left.\partial_y g_i(F-y'F_{y'})\right|_{x=x_i}\,.
\end{equation}
This equation is known as transversality condition
\cite{gelfand:fomin}. The curve $y(x)$ satisfying
\eref{eq:transversality} is said to be transversal to the curves
$g_i(x,y)=0$. 

\begin{figure}
  \centering
  \input{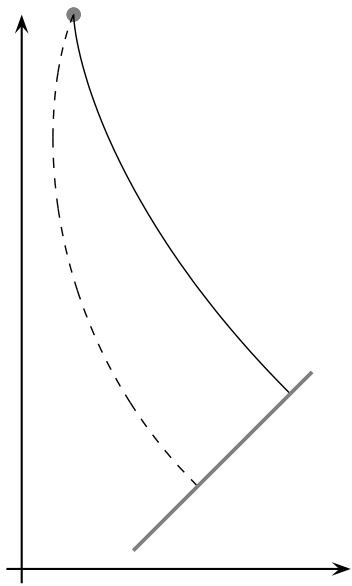}
  \caption{Brachistochrones for uniform gravitational field (solid)
    and $1/r$-potential (dashed) connecting a point with a straight
    line (gray). Both brachistochrones intersect the target curve
    orthogonally. The target curve is $y=x$, the starting point is
    $(1,2)$ The potentials are $v(x,y)=\sqrt{y_0-y}$ (uniform field)
    and $\sqrt{r^{-1}(x,y)- r^{-1}(x_0,y_0)}$ with
    $r(x,y)=\sqrt{x^2+y^2}$. For the numerical solution in the
    $1/r$-potential we used the integral representation of
    \cite{gemmer:nolan:umble}.}
  \label{fig:point-curve}
\end{figure}

Let us now consider variatonal problems of the brachistochrone type, i.e.
\begin{equation}
  \label{eq:brachis-functional}
  J[y] = \int\limits_{x_0}^{x_1} \frac{\sqrt{1+{y'}^2}}{v(x,y)}\,\D x\,,
\end{equation}
where $v(x,y)$ denotes the velocity of the particle at position $(x,y)$.
For problems like this, we get
\begin{displaymath}
  F_{y'} = \frac{y'}{1+{y'}^2}\,F \qquad F-y'F_{y'} = \frac{1}{1+{y'}^2}\,F\,,
\end{displaymath}
and the transversality conditions simplify to
\begin{equation}
  \label{eq:transverse-ortho}
  \partial_x g_i(x_i,y_i)\,y'(x_i) = \partial_y g_i(x_i,y_i)\,.
\end{equation}
The meaning of this equation becomes more transparent if we assume that
the curve $g_i(x,y)=0$ has an explicit representation $y=\phi_i(x)$. Then
\eref{eq:transverse-ortho} can be written as the orthogonality relation
\begin{displaymath}
  \phi'_i(x_i)\,y'(x_i) = -1\,.
\end{displaymath}
Hence for functionals of the form \eref{eq:brachis-functional},
transversality reduces to orthogonality. This proves Dr.~Grange's claim (B)
for arbitrary gravitational potentials (coded into $v(x,y)$) and arbitrary
target curves. Fig.~\ref{fig:point-curve} show brachistochrones
for the uniform field and the $1/r$-potential running between
a point and a straight line. The orthogonal intersection is clearly
visible.    

Note that orthogonality also applies for a varying start point.  For
$v(x,y)=\mathrm{const.}$ this means that the shortest \emph{distance}
between two curves is given by a straight line that intersects both
curves orthogonally.  Note also that the same argument (formal and
Dr.~Grange style) applies in higher dimensions: the brachistochrone
that connects a point with a surface is orthogonal to the surface.

\section{Visiting Dr.~Grange}

\begin{figure}
 \begin{center}
 \begin{pspicture}(0,0)(4,4)
   %\psgrid
   \psset{arrowsize=2pt 4}
   \psset{linecolor=black}
   %\psline{->}(0,0.2)(6,0.2)
   %\psline{->}(0.2,0)(0.2,4)
   %\listplot[plotstyle=curve,showpoints=false]{0.5 1 2.2 1.2 4 1.8} 
   \pscurve[linecolor=gray](0.0,1.4)(1.9,2.5)(2.5,4) 
   \pscurve[linecolor=gray](2,0)(3.2,0.8)(4,1.2) 
   \pscurve(2.2,3.0)(2.4,2)(3.2,0.8)
   \pscurve[linestyle=dashed](2,2.6)(2.2,1.6)(3,0.4)
   \psline{->}(2.2,3)(2,2.6)
   \psline{->}(3.2,0.8)(3,0.4)
    \end{pspicture}
    \caption{\label{fig:argc}  
      An infinitesimal translation of a
      sliding curve connecting two curves can save time if the slopes
      of the curves in the connected points are different.}
 \end{center}
\end{figure}
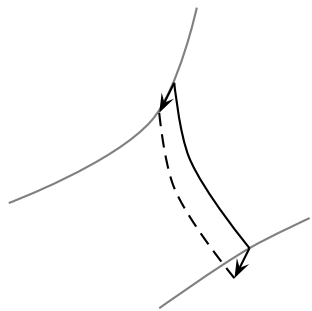

When we showed our analyis to Dr.~Grange, he was pleased that we
confirmed his heuristic reasoning. He pointed out, however, that our
treatment of the varying starting point does not properly reflect the
situation in the problem of the brachistochrone. Here the initial
velocity is usually zero, hence the integrand $F$ itself depends on
the starting point $(x_0,y_0)$ via
\begin{displaymath}
  F(x,y,y') = \frac{\sqrt{1+y'^2}}{\sqrt{2g(y_0-y)}}\,.
\end{displaymath}
So what is the brachistochrone that connects two given curves when we
start from rest?  Dr.~Grange remarked that his arguments (A) and (B)
are still valid, hence the type of the curve (cycloid) is fixed and
the orthogonality condition at the endpoint holds. For the starting
point Dr.~Grange gave the following argument:

\pagebreak%
\enlargethispage{\baselineskip}%

\begin{figure}
  \centering
  \input{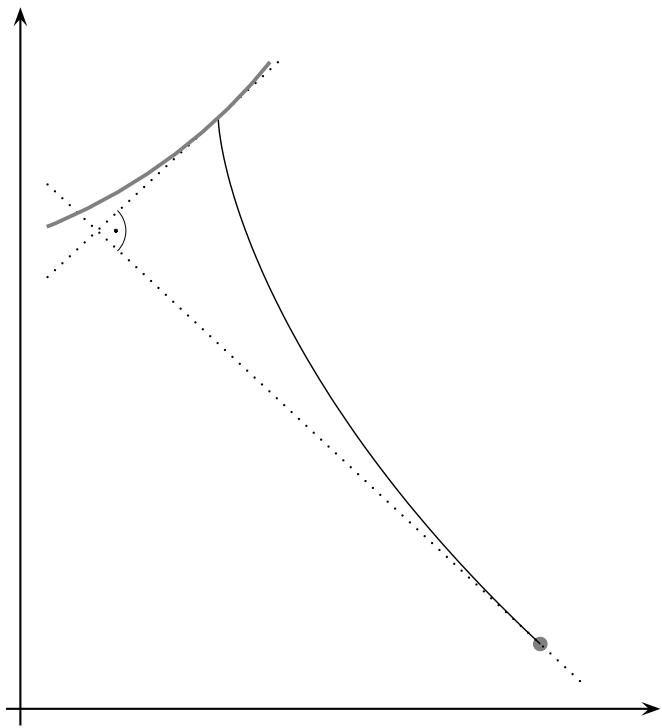}
  \caption{Brachistochrone in uniform gravitational field from a curve
  to a point. The slope of the brachistochrone in the target point 
  is orthogonal to the slope of the curve in the starting point.}
  \label{fig:curve-point}
\end{figure}

\begin{itemize}
\item[(C)] Consider a sliding curve that connects two points.
  If we shift the whole curve by a constant vector, the time to
  slide down the shifted curve is the same as before.  This follows
  from the translational symmetry of the uniform gravitational field.
  Let the sliding curve connect two curves $\mathcal{C}_0$ and
  $\mathcal{C}_1$. We move the whole sliding curve a little bit
  parallel or antiparallel to the tangent of $\mathcal{C}_0$ in the
  starting point. If the slope of $\mathcal{C}_1$ in the end
  point is different from the slope of the tangent, we can
  arrange that the shifted curve will intersect $\mathcal{C}_1$ before
  it reaches its endpoint (Fig.~\ref{fig:argc}). Hence we could save time
  by sliding down the shifted curve and, consequently, \emph{a brachistochrone
  that connects two curves intersects both curves in points of equal slope.} 
%   Consider a brachistochrone that runs from a point $P_0$ to
%   a point $P_1$. If we shift both points by the same vector, the
%   brachistochrone between the new points is simply the shifted curve,
%   and the time to slide it down is the same as before.  This follows
%   from the translational symmetry of the gravitational field. Let us
%   now assume that the brachistochrone that connects a curve
%   $\mathcal{C}_0$ with a curve $\mathcal{C}_1$ starts in
%   $P_0\in\mathcal{C}_0$ and ends in $P_1\in\mathcal{C}_1$. We move the
%   whole brachistochrone a little bit parallel or antiparallel to the
%   tangent of $\mathcal{C}_0$ in $P_0$. If the slope of $\mathcal{C}_1$
%   in $P_1$ is different from the slope of the tangent, we can arrange
%   that the shifted curve will intersect $\mathcal{C}_1$ before it
%   reaches its endpoint (Fig.~\ref{fig:argc}). Hence we could save time
%   by sliding down the shifted curve, which contradicts our assumption
%   that the original curve is a brachistochrone. We must conclude that
%   \emph{the brachistochrone between two given curves intersects both
%     curves in points of equal slope.}
\end{itemize}
Impressed by his intuition, we challenged Dr.~Grange by asking for the
brachistochrone with varying starting point but fixed endpoint.
Dr.~Grange smiled and gave the following argument:
\begin{itemize}
\item[(D)] Let us again move the whole brachistochrone a little bit
tangentially to the starting curve. The endpoint will move, too, and if
we imagine that its movement traces the tangent of a target curve with the same slope as the tangent of the starting curve, we know from (C) that
our brachistochrone is also the brachistochrone for this curve-to-curve
problem. From (B) we know that the brachistochrone is orthogonal to the
imaginary target curve. Hence (B) and (C) together imply that \emph{the slope
of the brachistochrone at the fixed endpoint is orthogonal to the slope of
the starting curve at the starting point (Fig.~\ref{fig:curve-point}).}
\end{itemize}
We thanked Dr.~Grange and headed back to our lab to prove his claims.

\section{Brachistochrones connecting curves}

We consider functionals where the
integrand depends explicetely on the starting point (like in the
Brachistochrone problem with zero initial velocity):
\begin{equation}
  \label{eq:functional-x0y0}
  J[y] = \int\limits_{x_0}^{x_1} F(x,y,y'; x_0,y_0)\,\D x\,.
\end{equation}
Repeating the arguments from Sec.~\ref{sec:varying-end} we get
\begin{equation}
   \label{eq:variation-x0-y0-2}
   \eqalign{
   \delta J = \left[ \big(F - y'F_{y'}\big)\delta x + F_{y'}\,\delta y\right]_0^1 
    + \int\limits_{x_0}^{x_1}\left(F_y -\diff{}{x} F_{y'}\right)\delta y\, \D x\cr
    \phantom{\delta ) = }+ \int\limits_{x_0}^{x_1}\left(F_{x_0}\delta x_0 + F_{y_0}\delta y_0\right) \D x\,.}
\end{equation}
Again we see that $\delta J=0$ implies the Euler equation
\eref{eq:euler}. The transversality conditions change, however, due
to the extra integral.  For the rest of this section we will assume
that $F$ depends on $x_0$ and $y_0$ only through $F(x-x_0,y-y_0,y')$
as for the brachistochrone in a uniform gravitational field. Then
we can evaluate the extra integrals using the Euler equation \eref{eq:euler}:
\begin{equation}
    \int\limits_{x_0}^{x_1} F_{y_0}\,\D x = -\int\limits_{x_0}^{x_1} F_{y}\,\D x 
  =-\int\limits_{x_0}^{x_1}\diff{}{x}F_{y'}\,\D x\nonumber
  = -\left[F_{y'}\right]_{x_0}^{x_1}
  \label{eq:Fy0}
\end{equation}
To evaluate the second integral, we use the Euler equation \eref{eq:euler} and
\begin{equation}
  \label{eq:Fx}
  \diff{}{x}F = F_x + F_y y' + F_{y'} y''\,.
\end{equation}
We find
\begin{eqnarray*}
     \int\limits_{x_0}^{x_1} F_{x_0}\,\D x &=& -\int\limits_{x_0}^{x_1} F_{x}\,\D x \\
   &=& \int\limits_{x_0}^{x_1} (F_y y' + F_{y'}y'')\,\D x - \Big[F\Big]_{x_0}^{x_1} \\
   &=& \int\limits_{x_0}^{x_1}\left[\left(\diff{}{x}F_{y'}\right) y' + F_{y'} y''\right]\,\D x - \Big[F\Big]_{x_0}^{x_1}\,.
\end{eqnarray*}
Partial integration provides us with
\begin{equation}
  \label{eq:Fx0}
  \int\limits_{x_0}^{x_1} F_{x_0}\,\D x = \left[F_{y'}\,y' - F\right]_{x_0}^{x_1}
\end{equation}
If we plug \eref{eq:Fy0} and \eref{eq:Fx0} into \eref{eq:variation-x0-y0-2},
we finally get
\begin{equation}
   \label{eq:variation-x0-y0-3}
   \eqalign{\delta J = \int\limits_{x_0}^{x_1}\left(F_y -\diff{}{x} F_{y'}\right)\delta y\, \D x + F_{y'}(x_1,y_1,y_1')(\delta y_1 - \delta y_0) \cr
    \phantom{\delta J = }+ \big[F(x_1,y_1,y_1') - y'(x_1) F_{y'}(x_1,y_1,y_1')\big](\delta x_1-\delta x_0)\,.}
\end{equation}
Now $\delta J=0$ implies $\delta y_1 = \delta y_0$ for $\delta x_1 =
\delta x_0$, which proves Dr.~Grange's claim (C) that a
brachistochrone between two curves meets both curves on points of
equal slope. If we set $\delta x_1 = \delta y_1 = 0$ we have
the problem of the brachistochrone from a curve to a fixed endpoint.
In this case the transversality conditions reduce to
\begin{equation}
  \label{eq:vary-startpoint}
  \fl 0 =  F_{y'}(x_1,y_1,y_1') \delta y_0 + \big[F(x_1,y_1,y_1') - y'(x_1) F_{y'}(x_1,y_1,y_1')\big] \delta x_0\,. 
\end{equation}
Comparison with \eref{eq:vary-endpoints} for $i=0$ proves Dr.~Grange's claim
(D).

\begin{figure}
  \centering
  \input{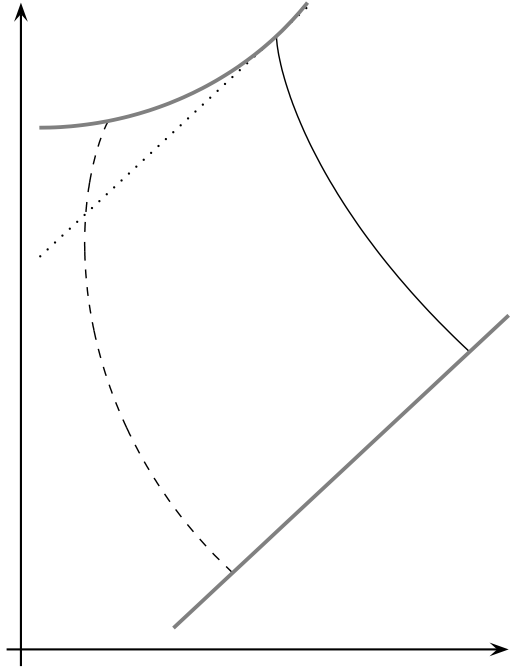}
  \caption{Brachistochrones for uniform gravitational field (solid)
    and $1/r$-potential (dashed) connecting a circle with a straight
    line (gray). Both brachistochrones intersect the target
    curve orthogonally, but only the brachistochrone in uniform field connects
    points of equal slope. The target curve is $y=x$, the circle is $(x-1)^2+(y-5/2)^2=1/4$.
    The potentials are the same as in Fig.~\ref{fig:point-curve}.}
  \label{fig:curve-curve}
\end{figure}

Note that the ``equal slope'' argument (C) relies on the translational
invariance of the uniform gravitational field (which we used in
our proof by assuming $F = F(x-x_0,y-y_0,y')$). It does not hold in
potentials like the $1/r$-potential (Fig.~\ref{fig:curve-curve}). 
The same is true for claim (D), allthough the
``orthogonality'' argument (B) applies for any potential.

\section{Conclusions}

We have seen that the generalized brachistochrone problem that allows
the endpoints to vary along given curves is not much different from
the classical version with fixed endpoints. In all cases the
brachistochrone is given by the solution of the same Euler equation.
The algebraic equations that fix the free parameters in this solution
depend on the boundary conditions, however. For brachistochrones from
a point to a curve, we have the orthogonality condition, i.e., the
brachistochrone intersects the target curve orthogonally.
Brachistochrones from a curve to a given point are determined by the
condition that the slope of the curve at the starting point is
orthogonal to the slope of the brachistochrone at the endpoint.
Finally, a brachistochrone from curve to curve is determined by the
orthogonality condition and the additional requirement that it
connects two points of equal slope. The two latter conditions only
hold for homogeneous gravitational potentials, whereas the
orthogonality condition holds for arbitrary potentials.

The fact that these results can be derived by simple, intuitive
arguments should encourage teachers to include the free boundary
version when they discuss the brachistochrone problem. The intuitive
solution of the free boundary problem helps to understand what a
variation actually means, a concept that many students do not easily
grasp. Working out the formal solution on the other hand is a good
exercise in using the machinerie of the variational calculus.

\ack{%
  We are grateful to Dr. L.A. Grange for sharing his insights with us.
}

\section*{References}

\bibliographystyle{unsrt} 
\bibliography{brachis,math}

% \newpage
% \section*{Figures}

% It is easy to include encapsulated postscript files (see Figure~\ref{fig:sine}). We can make figures bigger or smaller by scaling them. Figure~\ref{fig:lj} has been scaled by 80\%.
% Figures should be placed at the end of the manuscript and sent as separate files. It also is possible to include pdf files.

% \begin{figure}[h]
% \begin{center}
% \includegraphics{../figures/sine.eps}
% \caption{\label{fig:sine}Show me a sine.}
% \end{center}
% \end{figure}

\end{document}